\documentclass{article}
\usepackage{graphicx} 
\usepackage{cite}
\usepackage{amsmath,amssymb,amsfonts}
\usepackage{algorithmic}
\usepackage{graphicx}
\usepackage{textcomp}
\usepackage{chapterbib}  
\usepackage{caption}
\usepackage{subcaption}
\usepackage{array}
\usepackage{multirow}
\usepackage{caption}
\usepackage{subcaption}
\usepackage{booktabs}
\usepackage{multirow}
\usepackage{hyperref}
\usepackage[inkscapelatex=false]{svg}
\usepackage{pdfpages}
\usepackage{geometry}
 \usepackage{float}

\usepackage{authblk}
\title{\textbf{Towards Robust Cardiac Segmentation using Graph Convolutional Networks}}
\author[1]{Gilles Van De Vyver}
\author[2]{Sarina Thomas}
\author[3]{Guy Ben-Yosef}
\author[1]{Sindre Hellum Olaisen}
\author[1,4]{Håvard Dalen}
\author[1]{Lasse Løvstakken}
\author[1.5]{Erik Smistad}
\affil[1]{Norwegian University of Science and Technology, Trondheim, Norway} 
\affil[2]{University of Oslo, Oslo, Norway} 
\affil[3]{GE Research, Niskayuna, USA} 
\affil[4]{St. Olavs hospital \& Levanger hospital, Trondheim \& Levanger, Norway} 
\affil[5]{SINTEF Health, Trondheim, Norway} 
\date{}

\begin{document}

\maketitle
\begin{abstract}

Fully automatic cardiac segmentation can be a fast and reproducible method to extract clinical measurements from an echocardiography examination. The U-Net architecture is the current state-of-the-art deep learning architecture for medical segmentation and can segment cardiac structures in real-time with average errors comparable to inter-observer variability. However, this architecture still generates large outliers that are often anatomically incorrect. This work uses the concept of graph convolutional neural networks that predict the contour points of the structures of interest instead of labeling each pixel. We propose a graph architecture that uses two convolutional rings based on cardiac anatomy. While this architecture does not improve performance on classical measures like Dice score and Hausdorff distance, it does eliminate anatomical incorrect segmentations. 
Additionally, we propose to use the inter-model agreement of the U-Net and the graph network as a predictor of both the input and segmentation quality in real-time. The results show that from the 100 high agreement samples, 93 were in distribution, while from the 100 low agreement samples, only 7 were in distribution. Finally, this work contributes with an ablation study of the graph convolutional architecture on the publicly available CAMUS dataset and an evaluation of clinical measurements on the clinical HUNT4 dataset. 
 Source code is available online: \url{https://github.com/gillesvntnu/GCN_multistructure}

\begin{center}
    \textbf{Keywords}:
\end{center}
Cardiac segmentation, Ultrasound, Graph convectional network, U-Net, Robust segmentation
\end{abstract}

\section{Introduction}
\label{sec:introduction}

Cardiovascular diseases are the most common cause of death, accounting for one in three deaths in the United States \cite{benjamin2019heart}. Ultrasound imaging is the standard diagnostic tool for cardiology, as it is safe, inexpensive, real-time, and non-invasive. Accurate segmentation of the cardiac structures from ultrasound images is important for diagnosis, as it enables the extraction of standard clinical measurements such as left ventricular (LV) volume, ejection fraction (EF), and global longitudinal strain (GLS), which all provide important insights into the patient's cardiovascular health. In today's clinics, it is mainly done manually or semi-automatically. Clinical guidelines recommend that clinical measurements should be repeated over three cardiac cycles \cite{lang2015recommendations}. However, this is typically not performed since it takes too much time for the clinician. Automating segmentation would make repeating the measurements on multiple cardiac cycles trivial. 

With exponential growth in computing power and access to large amounts of digitized data, deep learning methods have become the most common approach for automatic medical image segmentation, showing average performance comparable to inter-observer variability \cite{smistad20172d,leclerc2019deep,Olaisen2023-fy}. Deep learning segmentation approaches usually directly predict a label for each pixel in the image. The U-Net architecture \cite{ronneberger2015u} is the most commonly used architecture for pixel-wise segmentation. U-Net is a fully convolutional neural network (CNN) that efficiently combines high and low-level features by integrating skip connections in an encoder-decoder structure. The output is a multi-channel segmentation map, with each channel corresponding to a structure. 

\begin{figure}
    \centering
    \includegraphics[width=0.6\textwidth]{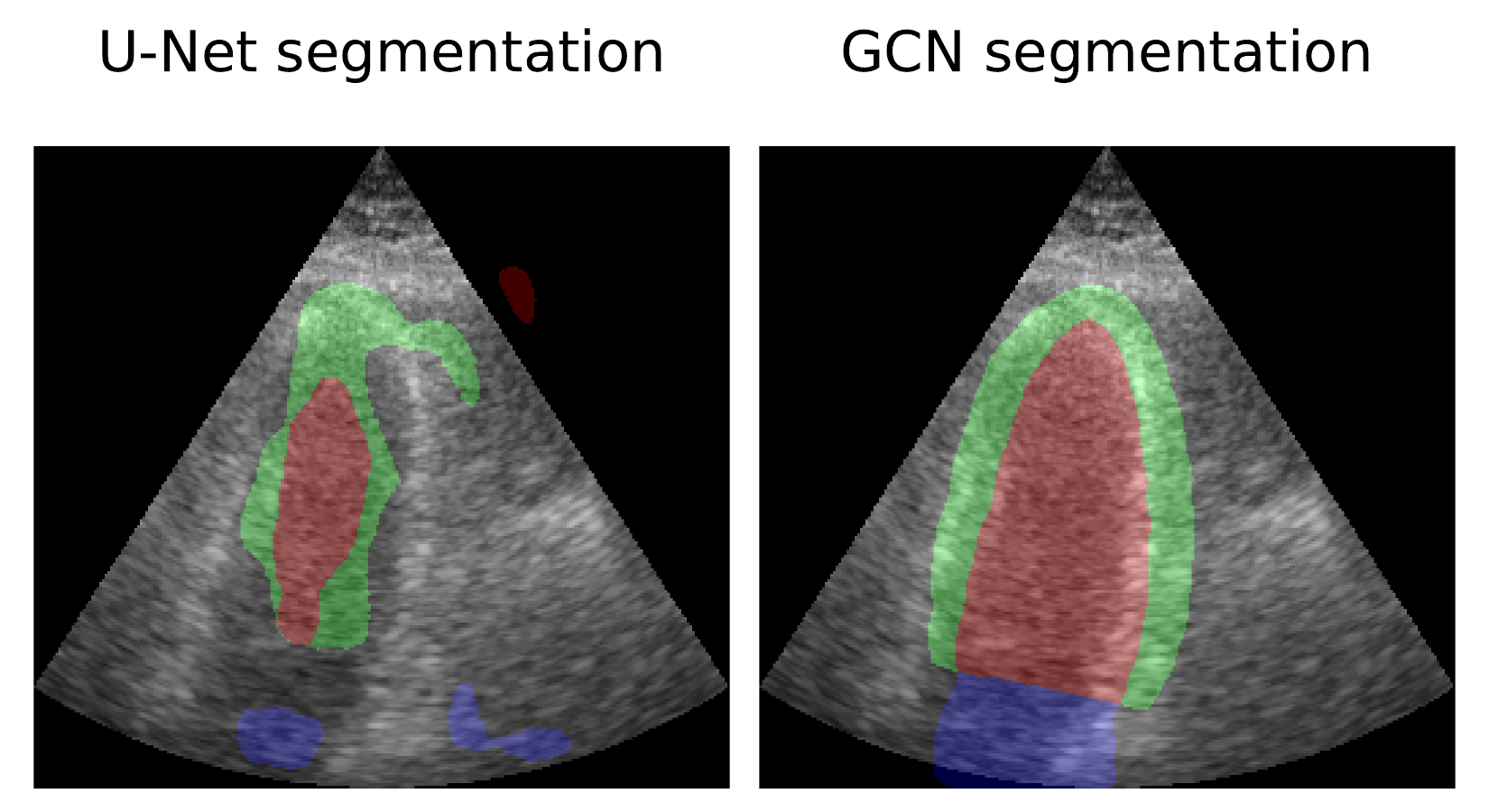}
    \caption{Though U-Nets can achieve a high average Dice accuracy on large datasets (> 0.94), they can still produce anatomical incorrect results as shown here to the left, with multiple atria disconnected from the LV and an incoherent myocardium around the LV, none of which are anatomically plausible. The anatomical correct output of the proposed graph convolutional network (GCN) is shown to the right.}
    \label{fig:anatomical_incorrect_example}
\end{figure}

Whereas the pixel-wise approach of U-Net gives it excellent accuracy on average, it has two fundamental issues:
\newpage 
\begin{enumerate}
    \item The U-Net has no anatomical constraints, which can result in outliers and anatomical incorrect segmentations, as shown in Fig.~\ref{fig:anatomical_incorrect_example}.
    \item The U-Net has no way of reporting when it is uncertain on how to segment an image, for instance in out-of-distribution and low image quality cases.
\end{enumerate} 
This lack of robustness hinders clinical use. Since the segmentation accuracy (e.g. Dice) of LV segmentation is on average already very high with U-Net methods, the goal of this paper is not to create a new network which improves these segmentation metrics, but instead investigate new methods for dealing with these two fundamental issues of anatomical incorrectness and detection of failing cases.

\subsection{Robust U-Net-based segmentation}

Several works have tried addressing the fundamental outlier problem of U-Net using shape priors. These works encourage the U-Net to produce anatomically valid shapes, either with a specialized loss function expressing anatomical correctness \cite{zotti2018convolutional}, an atlas prior \cite{dong2020deep}, closeness to a learned embedding \cite{oktay2017anatomically}, temporal consistency, \cite{smistad2021real} or with joint landmark detection \cite{duan2019automatic}. However, Painchaud et al. \cite{painchaud2020cardiac} showed that methods using soft constraints on pixel labeling methods still produce anatomically incorrect outputs. Their work proposed a post-processing approach that refines the output of the U-Net to the closest anatomically valid shape \cite{painchaud2020cardiac} in a large latent space. The architecture consists of an autoencoder trained on ground truth segmentations and generates a latent space of valid anatomically correct segmentations. During inference, the output of the U-Net-based model is then fed to the encoder producing a latent representation. A nearest neighbor search is performed in the latent space and fed to the decoder to produce an anatomically corrected output. This method shows excellent results but is too slow for real-time applications. 
However, the work introduces a real-time variant that works as a denoising autoencoder. In this work, we compare our results to this version on the CAMUS dataset \cite{leclerc2019deep}.

\subsection{Graph Neural Networks}

Graph Convolutional Networks (GCN) address the problem of segmentation in a fundamentally different way. Instead of predicting a pixel-wise segmentation map, these models predict the contour of the segmentation as a graph. The contours are sampled into keypoints which form the nodes of a graph. Previous works have shown the effectiveness of this method for MRI segmentation \cite{tian2020graph} and X-ray \cite{gaggion2022improving}. In our previous work, we demonstrated the feasibility of GCNs for ultrasound segmentation \cite{thomas2022light} using the EchoNet dataset \cite{ouyang2019echonet}. However, this work was limited to single-structure segmentation (LV endocardium) on a single view (apical four-chamber) and lacked a thorough evaluation. 

\subsection{Contributions}
In this work, we extend our work on the GCN approach for cardiac ultrasound segmentation with a focus on the two fundamental issues highlighted in the introduction; anatomical correctness and detection of failing cases.
\begin{enumerate}
  \item \textbf{Multi-structure segmentation:} We explore methods for extending the graph to segment multiple structures, such as the LV epicardium and left atrium (LA), enabling the segmentation to be used for more measurements such as strain. We propose a clinically motivated design and show that this eliminates anatomically incorrect shapes in the segmentation output of the network. We perform an ablation study of the GCN architecture on the public CAMUS dataset and compare it with U-Net. Additionally, we evaluate the final models on the clinical HUNT4 dataset.
  
  \item \textbf{Segmentation quality predictor:} We combine the outputs of the GCN and the U-Net and show that this can be used to detect out-of-distribution and unsuitable low image quality cases, resulting in bad segmentation output.
  \item \textbf{Open-source framework and demo:} We make our code publicly available. We provide two parts: the full PyTorch framework to reproduce the results (\url{https://github.com/gillesvntnu/GCN_multistructure}) and the C++ code to run the real-time demo application described later in this article (\url{https://github.com/gillesvntnu/GCN_UNET_agreement_demo}).
\end{enumerate}

\section{Methods}
\label{sec:methods}

\subsection{U-Net baseline models} \label{sec: baseline unets}
We compare the GCN with two U-Net architectures. U-Net 1 \cite{leclerc2019deep}, a fixed architecture optimised for speed, and nnU-Net V2 \cite{nnUNetV2,isensee2021nnu},  a self-adapting framework optimised for accuracy. For U-Net 1, we use the same architecture and training procedure as in \cite{leclerc2019deep}, but with additional augmentations as listed in Table \ref{table: characteristics unets}. These are the same augmentations as in \cite{thomas2022light}. The nnU-Net is used out of the box using the default configuration, but without the final ensemble step \cite{nnUNetV2}\cite{isensee2021nnu}. For both U-Nets, the input images are resized to 256x256 pixels as a pre-processing step. 

\begin{table*}
\tiny
  \centering
  \caption{Characteristics of the U-Net baselines. U-Net 1 and nnU-Net are the same as in \cite{leclerc2019deep} and \cite{nnUNetV2} respectively. The "number of channels" column indicates the number of channels at the first, bottom, and last convolution of the U-Net. Table \ref{table: runtime} shows the inference time and number of parameters for each network.}
  \begin{tabular}{m{33pt}m{43pt}m{33pt}m{43pt}m{43pt}m{17pt}m{30pt}m{34pt}m{30pt}m{80pt}}
    \toprule
    \textbf{Architecture} & \raggedright \textbf{Number of channels} & \raggedright \textbf{Lowest resolution} & \textbf{Upsampling scheme} & \textbf{Normalization scheme} & \textbf{Batch Size} & \textbf{Learning Rate} & \textbf{Scheduler} & \textbf{Loss function} & \textbf{Augmentations}\\ \midrule
    U-Net 1 & 32 $\downarrow$ 128 $\uparrow$ 16 & 8*8 & 2*2 repeats & None & 32 & 1e-3 & None & Dice &  Rotations, scaling, cropping, brightness adjustments and mirroring \\
    \midrule
     nnU-Net & 32 $\downarrow$ 512 $\uparrow$ 32 & 4*4 & Deconvolutions & InstanceNorm & 49 & 1e-2 & Polynomial & \raggedright Dice \& cross-entropy & Rotations, scaling, Gaussian noise, Gaussian blur, brightness, contrast, simulation of low resolution, gamma correction and mirroring
     \\
     \bottomrule
    \label{table: characteristics unets}

  \end{tabular}
\end{table*}

\subsection{General structure of the GCN}

\begin{figure*}
\centering
  \centering
  \includegraphics[trim={0cm 0cm 0cm 0cm},clip,width=1\textwidth]{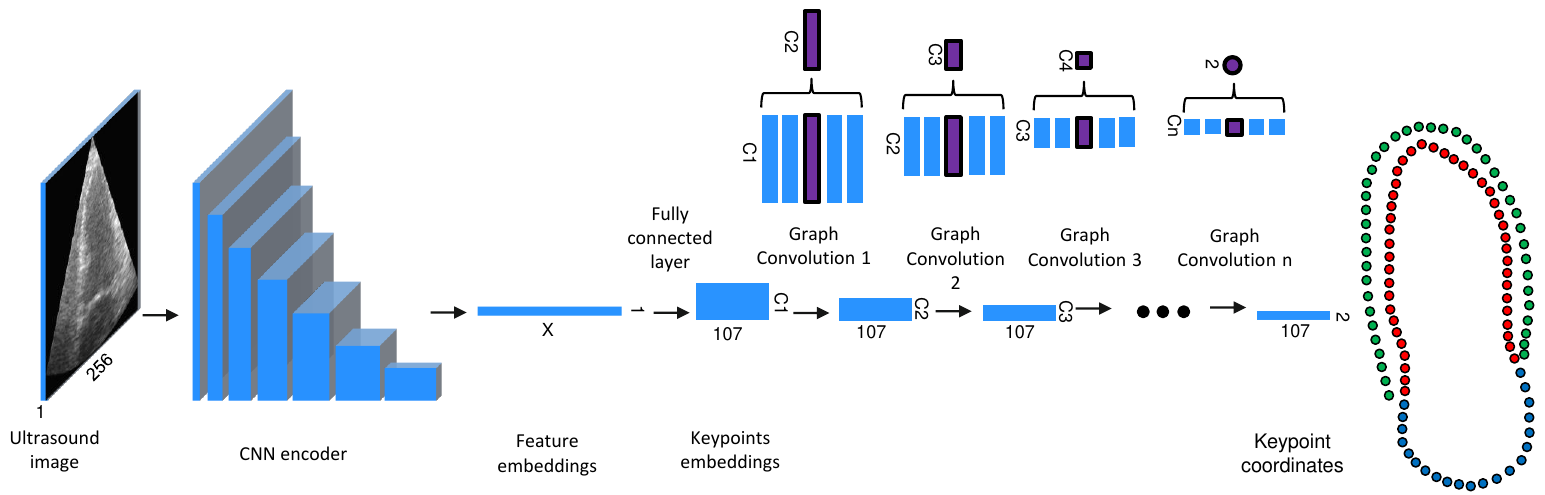}
  \caption{The architecture of the GCN. The CNN encoder transforms the input ultrasound image of width $W$ and height $H$ to an embedded vector of size $X$. A dense layer transforms this embedding to an embedding in keypoint space, with $107$ keypoints and $C1$ channels. The decoder consists of a sequence of graph convolutions over these keypoint embeddings.
  The final outputs are the 2D coordinates of the keypoints in the image. }
  \label{fig: basic GCN}
\end{figure*}

The general structure of a GCN for cardiac segmentation \cite{thomas2022light} consists of a CNN encoder pre-trained on ImageNet \cite{deng2009imagenet} that acts as a feature extractor and a graph convolutional decoder that reconstructs the keypoints of the contours of the cardiac structures. Fig.~\ref{fig: basic GCN} shows a general overview of the GCN architecture. The encoder transforms a grayscale ultrasound image of $\mathbb{N}^{H \times W}$ pixel values to a vector embedding of $\mathbb{R}^{X}$ values, where $X$ is the feature embedding size of the last layer of the encoder. A dense layer transforms these embeddings to a representation in the keypoint space of size $\mathbb{R}^{n \times C1}$, where $n$ is the number of keypoints and C1 is the number of output channels in the first layer of the GCN decoder. The decoder layers perform graph convolutions in the keypoint space and gradually decrease the number of channels until the final output with dimensions $\mathbb{R}^{n \times 2}$, representing the relative pixel coordinates for each keypoint in the image. The size of the intermediate channels of the keypoint embeddings $C_{i}$ is a tunable parameter. 


Every graph convolutional layer consists of a linear layer followed by an activation function. It takes the embeddings of adjacent keypoints as input to update the embedding of each keypoint. In the ring of LV keypoints $p_{1..n}$, the embeddings of keypoints $p_{i-w..i+w}$, are the input to update the embeddings of keypoint $p_i$, where $w$ is the receptive field of the GCN decoder. The same weights are reused at each keypoint, creating an inductive bias of locality. Section \ref{subsec: multi seg} describes how to expand this architecture to segment multiple structures simultaneously.

\subsection{Keypoint extraction} \label{subsec: data preproc}
\begin{figure}
\centering
  \centering
  \includegraphics[width = 0.5\linewidth]{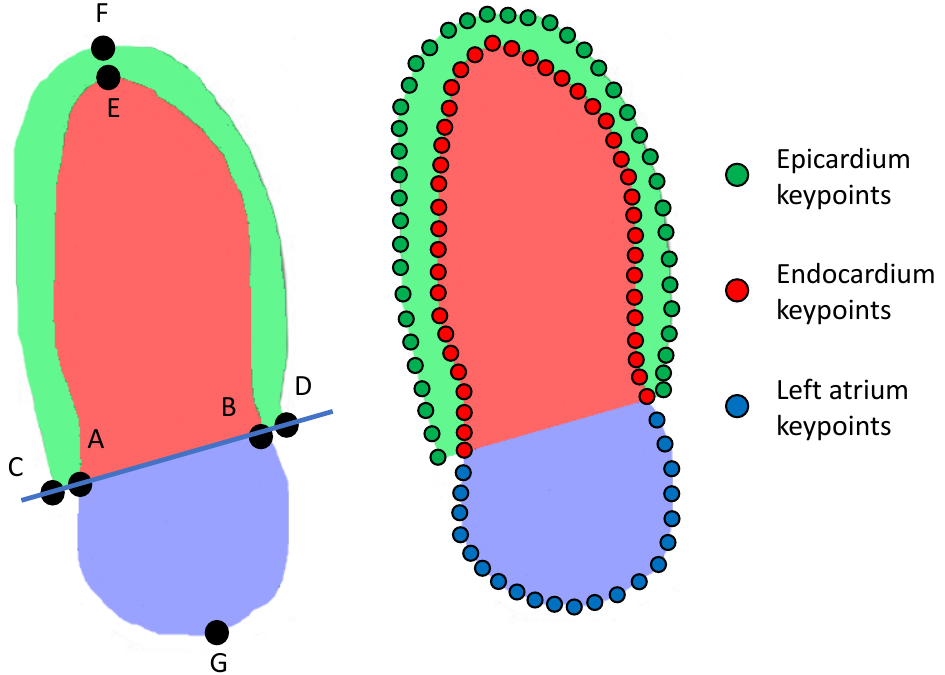}
  \caption{Schematic diagram showing the preprocessing to transform pixel labels to keypoints positions. A and B are the base points of the LV. C and D are the base points of the MYO. E, F, and G are the apexes of the LV, MYO, and LA respectively.}
  \label{fig: preprocess}
\end{figure}

The keypoints are extracted in a standardized way from the annotations that are available as pixel-wise labels of the left ventricle (LV), left atrium (LA), and myocardium (MYO). 
The following algorithm extracts anatomical landmarks from the annotations and uniformly samples the contours between these landmarks to fill in the remaining keypoints, as shown in Fig.~\ref{fig: preprocess}. The anatomical landmarks A-G have a unique physical meaning/location, while the rest of the keypoints are sampled uniformly on the contour between these anatomical landmarks as described in the algorithm below:

\begin{enumerate}
  \item Extract the annulus points, the corner points where the MYO meets the LA. These are points A and B in Fig.~\ref{fig: preprocess}. Connecting these points gives the base line.
  \item Extract the base points of the MYO by extending the base line and selecting the furthest point with the MYO label on the base line. These are points C and D in Fig.~ \ref{fig: preprocess}
  \item Extract the apexes of the LV, MYO, and LA, defined as the furthest points from the base line with the corresponding label. These are points E,F, and G respectively in Fig.~\ref{fig: preprocess}
  \item Sample the contour of each structure to complete the keypoint extraction. Sample the endocardium contour line between E and A in Fig.~\ref{fig: preprocess} for \textit{n} equidistant points and do the same for the endocardium contour line between E and B, with \textit{n} a model parameter. The total number of endocardium points is then \textit{2n + 3}, consisting of the sampled points together with A,B, and E. Follow the same procedure for the epicardium, using F,C, and D as corner points. This results in another \textit{2n + 3} keypoints for the epicardium, as the number of endo- and epicardium keypoints needs to be equal in our architecture. Finally, sample the LA border between G and A for \textit{m} equidistant points and do the same between G and B, with \textit{m} another model parameter. The total number of LA keypoints is then \textit{2m + 1}, consisting of the sampled points together with G. We do not add A and B to the LA points because these points already belong to the endocardium keypoints. In this work, we choose \textit{n=20} and \textit{m=10}, resulting in a total of 107 keypoints. The right side of Fig.~\ref{fig: preprocess} shows the final set of keypoints. 
\end{enumerate}



\subsection{Multi-structure segmentation} \label{subsec: multi seg}

The graph convolutions of the GCN are performed over the  keypoints in the spatial neighborhood in the graph. Since the original approach only included the endocardium, the graph convolutions had to be adapted for the additional structures; the epicardium and the LA. 

\begin{figure}
     \centering
     \begin{subfigure}[b]{0.25\textwidth}
         \centering
\includegraphics[width=1\textwidth]{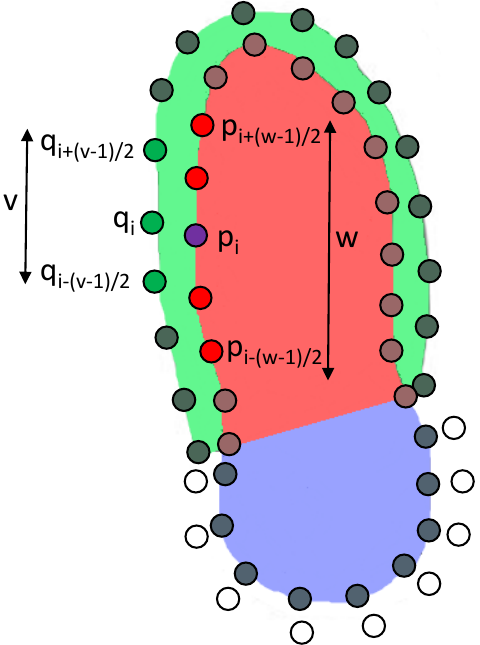}
         \caption{Inner ring convolution}
         \label{fig:inner conv}
     \end{subfigure}
     \hspace{0.1cm}
     \begin{subfigure}[b]{0.25\textwidth}
         \centering \includegraphics[width=1\textwidth]{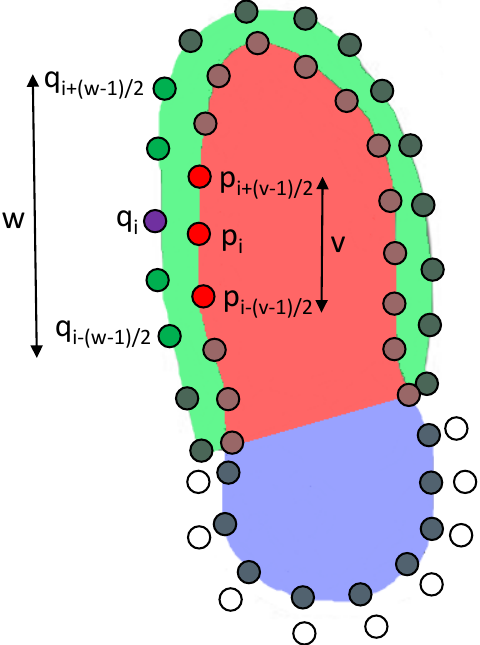}
         \caption{Outer ring convolution}
         \label{fig:outer conv}
     \end{subfigure}
        \caption{Schematic diagram showing the multi-structure convolution. $q_{1..n}$ are zero-padded epicardium keypoints, $p_{1..n}$ are endocardium and left atrium keypoints. The highlighted points are used as input to update the embedding of the purple keypoint, with w the primary receptive field and v the secondary receptive field.For illustrative purposes, the diagram does not show the actual number of keypoints used in this work.}
        \label{fig: multi_conv}
    \hspace{1cm}
\end{figure}

To include the LA keypoints, the ring of $n$ endocardium keypoints $p_{1..n}$ from the original GCN was extended to a ring of keypoints $p_{1..n+m}$, where $m$ is the number of LA keypoints. The endocardium and LA keypoints together form the keypoints in the inner ring. In our implementation, $n=43$ and $m=21$ keypoints are used. For the epicardium keypoints, a second ring of keypoints $q_{1..n}$ was added, which are zero-padded to form a ring of $q_{1..m+n}$ outer keypoints. In the graph convolutional layers of the decoder, the embeddings of inner keypoints  $p_{i-(w-1)/2..i+(w-1)/2}$ and outer keypoints $q_{i-(v-1)/2..i+(v-1)/2}$ are the input to produce the embeddings of each inner ring keypoint $p_i$, where $w$ is the primary receptive field and $v$ is the secondary receptive field of the GCN decoder. The weights are reused at each keypoint, creating an inductive bias of locality. Fig.~\ref{fig:inner conv} visualizes the inner convolution schematically. For each epicardium keypoint $q_i$, the embeddings of the outer keypoints $q_{i-(w-1)/2..i+(w-1)/2}$ and inner keypoints $p_{i-(v-1)/2..i+(v-1)/2}$ are the input to the convolutional layer. Also, for the epicardium the weights are reused at each keypoint, but the weights of the inner ring are different than the weights of the outer ring. Fig.~\ref{fig:outer conv} visualizes the outer convolution schematically.


\subsection{Displacement method data representation}
An alternative way to represent the epicardium keypoints is to use the thickness of the MYO at each of the endocardium keypoints. The position of each epicardium keypoint is then defined by the normal distance relative to the corresponding keypoint of the endocardium. The normal direction is calculated using the two neighboring keypoints. Only the final layer of the graph decoder needs to be adjusted to the new output format. The displacement representation prevents the network from producing erroneous segmentations where the epicardium keypoints are inside the endocardium keypoints.

\subsection{GCN training procedure}

The training procedure of the GCN is the same as in \cite{thomas2022light}. It uses the Adam \cite{kingma2014adam} optimizer using a learning rate of 1e-5. The model is trained for 5000 epochs and the model weights with the best validation accuracy are retained. The training data is resized to 256x256 pixels and augmented by using rotations, scaling, cropping, brightness adjustments, and mirroring. The loss function is the sum of the Euclidean distances of all predicted keypoints plus the mean absolute error of the displacements, if applicable.

\subsection{Combining U-Net and GCN}

\subsubsection{GCN - U-Net cascade model} \label{subsec: cascade}
\begin{figure}
\centering
  \centering
  \includegraphics[width = 0.7\linewidth]{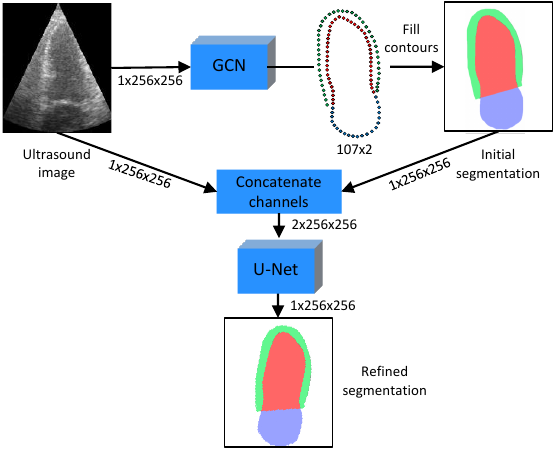}
  \caption{Graphical representation of the GCN - U-Net cascade. The cascade model concatenates the initial segmentation produced by the GCN with the original ultrasound input and feeds this to the U-Net to produce the final, refined segmentation. The segmentations shown are only illustrative.}
  \label{fig: cascade}
\end{figure}

 We explore a cascade network of GCN with displacement method followed by U-Net. The idea is that the GCN with displacement method will produce an anatomically correct initial shape, which the U-Net can refine to a more accurate pixel-wise segmentation. In the first stage, the GCN with displacement method is trained as before. After training, the GCN performs inference on the training set, and the resulting keypoints are transformed into pixel-wise segmentation outputs. In the second stage, a U-Net is trained with the concatenation of the grayscale ultrasound image and the segmentation output of the GCN as input. The only change to the U-Net architecture is that the first layer takes an extra input channel. We use the same U-Net architectures as described in section \ref{sec: baseline unets}. For U-Net 1, the input segmentation channel is additionally augmented separately with rotation, scaling, and translation to avoid the U-Net from fully relying on the segmentation outputs of the GCN. 
For nnU-Net, the GCN segmentation serves as an extra input channel and the framework is used out of the box without custom augmentations. Fig.~\ref{fig: cascade} shows a graphical representation of the cascade model.

\subsubsection{GCN - U-Net ensemble}
For estimating ejection fraction using the GCN - U-Net ensemble, both networks run in parallel and construct their segmentations, leading to two different ejection fraction estimations. The average of these two measurements is the output of the ensemble.

\subsection{Inter-model agreement as a quality predictor}

The second goal of this work was to develop a method to detect when the segmentation fails.When the GCN and U-Net generate segmentations in parallel, the agreement of the two models in terms of Dice agreement between the two segmentation outputs contains additional information. The idea is that if the two methods, each with their own data representation, generate similar segmentations, there is a higher probability of the segmentations being correct. On the other hand, when one of the methods makes a large error, for example, due to an out-of-distribution case, it is unlikely the other method will make the same mistake because of their fundamental differences in data representation, as demonstrated in Fig.~\ref{fig:anatomical_incorrect_example}. 
Thus we propose to use the inter-model agreement of these two different segmentation architectures to detect failing segmentation cases.


\section{Experimental setup and Results}
\label{sec: experimental setup and results}

\subsection{Datasets}

\subsubsection{CAMUS}

The CAMUS dataset is a publicly available dataset of 500 patients including apical 2 chamber (A2C) and apical 4 chamber (A4C) views obtained from a GE Vivid E95 ultrasound scanner, equalling 2000 image annotation pairs \cite{leclerc2019deep}. The annotations are available as pixel-wise labels of the left ventricle (LV), left atrium (LA), and myocardium (MYO), split into 10 folds for cross-validation. 

\subsubsection{HUNT4}

The \textit{Helse Undersøkelsen i Nord-Trøndelag} ultrasound dataset (HUNT4Echo) is a large-scale clinical dataset of 2462 patient examinations of LV-focused A2C and A4C recordings used for clinical measurements \cite{Olaisen2023-fy}. Each recording contains 3 cardiac cycles. A fraction of 311 patient exams, the training set, contains single frame segmentation annotations in both ED and ES as pixel-wise labels of the LV, LA, and MYO. For 1913 patient exams, there is a reference value of the biplane LV volumes in end-diastole (ED) and end-systole (ES), obtained manually using the clinically approved EchoPAC software (GE HealthCare).

\subsubsection{Differences}

Due to differences in annotation conventions and scanning approaches, the HUNT4 and CAMUS datasets can not be used jointly. For instance, the myocardium is consistently annotated to be much thicker in CAMUS than in HUNT4. Although both datasets contain cardiac images of the same clinical views, the images in the HUNT4 dataset are consistently LV-focused, which is not the case for CAMUS. Whereas we consider the HUNT4 annotations to be more clinically correct, we have used and included results for the CAMUS dataset as well in our evaluation since it is publicly available whereas HUNT4 is not.

\subsection{Ablation study}
The ablation study uses the first subgroup of the CAMUS dataset, meaning it tests on the first cross-validation split, validates on the second and uses the remaining eight splits for training. The first experiment varies the decoder while keeping the CNN encoder fixed to MobileNet-v2 \cite{sandler2018mobilenetv2}. The second experiment varies the encoder.

Table \ref{table: ablation study} summarizes the results of the first part of the ablation study that focuses on varying the complexity of the decoder. The channels are the embedding sizes of the decoder layers, corresponding to $C_{1..n}$ in Fig.~\ref{fig: basic GCN}. The secondary receptive field is the number of keypoints used from the second ring, as described in section \ref{subsec: multi seg}. The primary receptive field is kept constant at $w=n+m$, 
the total number of inner keypoints, as in the original GCN. When there are no channels, the decoder is replaced by a single dense layer, converting the feature embeddings from the encoder directly to keypoint coordinates. The difference in overall Dice score between the best and worst performing variation is not statistically significant, $p=0.77$ and $p=0.58$ with and without displacement method respectively, using the Wilcoxon signed-rank test\cite{woolson2007wilcoxon}. The remainder of this work uses the GCN version with two decoder layers and a secondary receptive field of one.

The second part of the ablation study varies the CNN encoder. Table \ref{table: ablation study part 3} compares the Dice scores of the network with MobileNet-v2 \cite{sandler2018mobilenetv2} and ResNet-50 \cite{he2016deep} as encoder. The difference in overall Dice score is statistically significant ($p<0.05$) for the version with displacement method and not significant ($p=0.41$) for the version without displacement method. While slightly more accurate, the ResNet-50 encoder makes the model slower and an order of magnitude larger, as shown in Table \ref{table: runtime}. Therefore, the remainder of this work uses MobileNet-v2 as encoder.

\begin{table} 
\begin{center}
\scriptsize
\caption{Ablation study GCN part one.
The difference in Dice score between the best and worst performing variation is not statistically significant with or without displacement method.
}
\label{table: ablation study}
\begin{tabular}{m{70pt}|m{80pt}m{110pt}m{50pt}m{50pt}m{50pt}}
\toprule
\textbf{Decoder} & \textbf{Channels} & \textbf{Secondary receptive field} & \textbf{Dice LV} & \textbf{Dice MYO} & \textbf{Dice LA} \\
\midrule
& 48,32,32,16,16,8,8,4 & 11 & .910$\pm$.07 & .813$\pm$.15 & \textbf{.867}$\pm$.15 \\
\multirow{2}{*}{GCN} & 32,16,8,4 & 5 & .910$\pm$.07 & .811$\pm$.14 & .862$\pm$.09  \\
& 8,4 & 1 & \textbf{.914}$\pm$.05 & \textbf{.819}$\pm$.13 & .861$\pm$.09 \\
& None & n.a. & .911$\pm$.07 & .811$\pm$.14 & .860$\pm$.11\\
\hline
\multirow{4}{2cm}{GCN with displacement method}& 48,32,32,16,16,8,8,4 & 11 & .908$\pm$.08 & .810$\pm$.11 & .863$\pm$.09\\
  & 32,16,8,4 & 5 & .909$\pm$.07 & .809$\pm$.12 &.863$\pm$.09\\
& 8,4 & 1 & \textbf{.911}$\pm$.06 & \textbf{.812}$\pm$.11 & \textbf{.864}$\pm$.10\\
& None & n.a. & .907$\pm$.08 & .809$\pm$.12 & .856$\pm$.11\\
\bottomrule
\end{tabular}
\label{tab1}
\end{center}
\end{table}

\begin{table} 
\begin{center}
\scriptsize
\caption{Ablation study GCN part two. 
The difference in Dice score between the two decoders is statistically significant.
}
\label{table: ablation study part 3}
\begin{tabular}{m{150pt}|m{120pt}m{50pt}m{50pt}m{50pt}}
\toprule
\textbf{Decoder} & \textbf{CNN encoder} & \textbf{Dice LV} & \textbf{Dice MYO} & \textbf{Dice LA} \\
\midrule
\multirow{2}{*}{GCN} & MobileNet-v2 \cite{sandler2018mobilenetv2} & \textbf{.914}$\pm$.05 & .819$\pm$.13 & .861$\pm$.09\\
 & ResNet-50 \cite{he2016deep} &  .912$\pm$.06 &  \textbf{.824}$\pm$.11 &  \textbf{.874}$\pm$.08  \\
\hline
\multirow{2}{4cm}{GCN with displacement method}& MobileNet-v2 \cite{sandler2018mobilenetv2} &  .911$\pm$.07&  .802$\pm$.16&  .862$\pm$.09 \\
 & ResNet-50 \cite{he2016deep} & \textbf{.911}$\pm$.06 & \textbf{.812}$\pm$.11 & \textbf{.864}$\pm$.10\\
\bottomrule
\end{tabular}
\label{tab1}
\end{center}
\end{table}


\subsection{Anatomical correctness and segmentation accuracy}

\begin{table}
\scriptsize
\begin{center}
\caption{{Anatomical correctness and segmentation accuracy (Dice and Hausdorff) comparison} with SOTA on all cross-validation splits of CAMUS.}

\label{table: comp CAMUS}
\begin{tabular}{m{150pt}m{60pt}m{60pt}m{60pt}}\toprule
\textbf{Method} & \textbf{\#Anatomical}\newline\textbf{incorrect}$^*$ & \textbf{Dice score}$^{\dagger}$ & \textbf{Hausdorff}\newline\textbf{distance (mm)}$^{\dagger}$\\
\midrule
rVAE post processing \cite{painchaud2020cardiac}& 5-22$^{\ddagger}$& .916-.923$ \pm$n.a.$^{\ddagger}$& 5.7-6.2$ \pm$n.a.$^{\ddagger}$\\
nnU-Net \cite{isensee2021nnu}& 25 & \textbf{.949$\pm$.03}& \textbf{4.56$\pm$2.6} \\
U-Net 1 \cite{leclerc2019deep}  & 68& .943$\pm$.04 & 5.43$\pm$4.7\\
GCN & 12 & .936$\pm$.04&5.20$\pm$2.3 \\
GCN displacement & \textbf{0}& .932$\pm$.05 & 5.96$\pm$2.8\\
GCN displacement - U-Net1 cascade  & 32& .943$\pm$.03 & 5.17$\pm$2.9 \\
GCN displacement - nnU-Net cascade  & \textbf{0} & .938$\pm$.03 & 5.15$\pm$2.6\\
\bottomrule
\multicolumn{4}{p{380pt}}{
$^*$Anatomical correctness was measured with the same criteria as Painchaud et al. \cite{painchaud2020cardiac}.\newline
$^{\dagger}$The Dice score is the average of the $LV_{Endo}$ and $LV_{Epi}$ Dice scores \cite{leclerc2019deep}. The Hausdorff distance is the average of the $LV_{Endo}$ and $LV_{Epi}$ \cite{leclerc2019deep}.\newline
$^{\ddagger}$Painchaud et al. \cite{painchaud2020cardiac} report several values from different U-Nets. Here, the extreme values are reported. 
}
\end{tabular}
\label{tab1}
\end{center}
\end{table}

As the purpose of this work was to create a more anatomical correct segmentation method, an experiment was conducted to measure the number of anatomical incorrect segmentation. For this purpose we use the publicly available CAMUS dataset and the same criteria for anatomical correctness as Painchaud et al. \cite{painchaud2020cardiac}.
In addition, we calculate the Dice and Hausdorff distance to demonstrate the trade-off between anatomical correctness and segmentation accuracy.
We performed this experiment using our GCN architectures and other state-of-the-art (SOTA) cardiac segmentation methods for comparison, and the results are summarized in Table \ref{table: comp CAMUS}.
As this study focuses on real-time applications, we compare them with the real-time version of the post-processing method proposed by Painchaud et al., the robust Variational AutoEncoder (rVAE) \cite{painchaud2020cardiac}. Furthermore, we compare to U-Net 1 \cite{leclerc2019deep} and nnU-Net \cite{isensee2021nnu}. To be able to compare to previous work, the Dice score is calculated in the same way as in the work of Painchaud et al. \cite{painchaud2020cardiac} where the myocardium is added to the LV lumen to create an "epicardium" region which results in overall higher Dice scores. The Dice score and Hausdorff distance are calculated for the LV by itself ($LV_{endo}$ in \cite{leclerc2019deep}) and the LV and MYO together ($LV_{epi}$ in \cite{leclerc2019deep}). The final result is the average of these two values. 
With $p<0.05$ using the Wilcoxon signed-rank test\cite{woolson2007wilcoxon}, the Dice score of nnU-Net is significantly higher than that of the GCN, regardless of the presence of the displacement method.

\begin{figure}
\centering
  \centering
  \includegraphics[width = 0.6\linewidth]{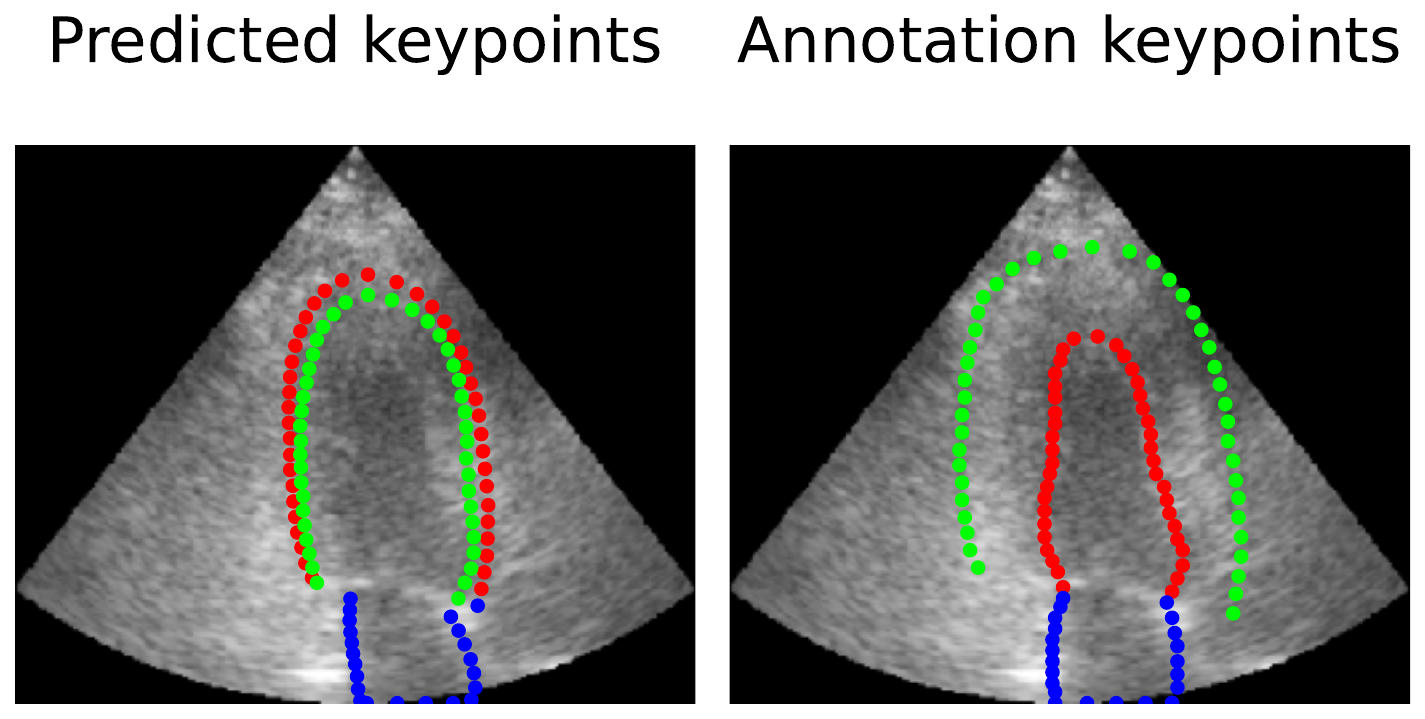}
  \caption{Anatomically incorrect case for the GCN without displacement method. The predicted keypoints of the endo- and epicardium are flipped and partly overlapping. The displacement method eliminates these errors.}
  \label{fig: ring_error}
\end{figure}

\begin{figure*}
\centering
\begin{subfigure}[b]{0.8\textwidth}
   \includegraphics[trim={0cm 0cm 0cm 0cm},clip,width=1\linewidth]{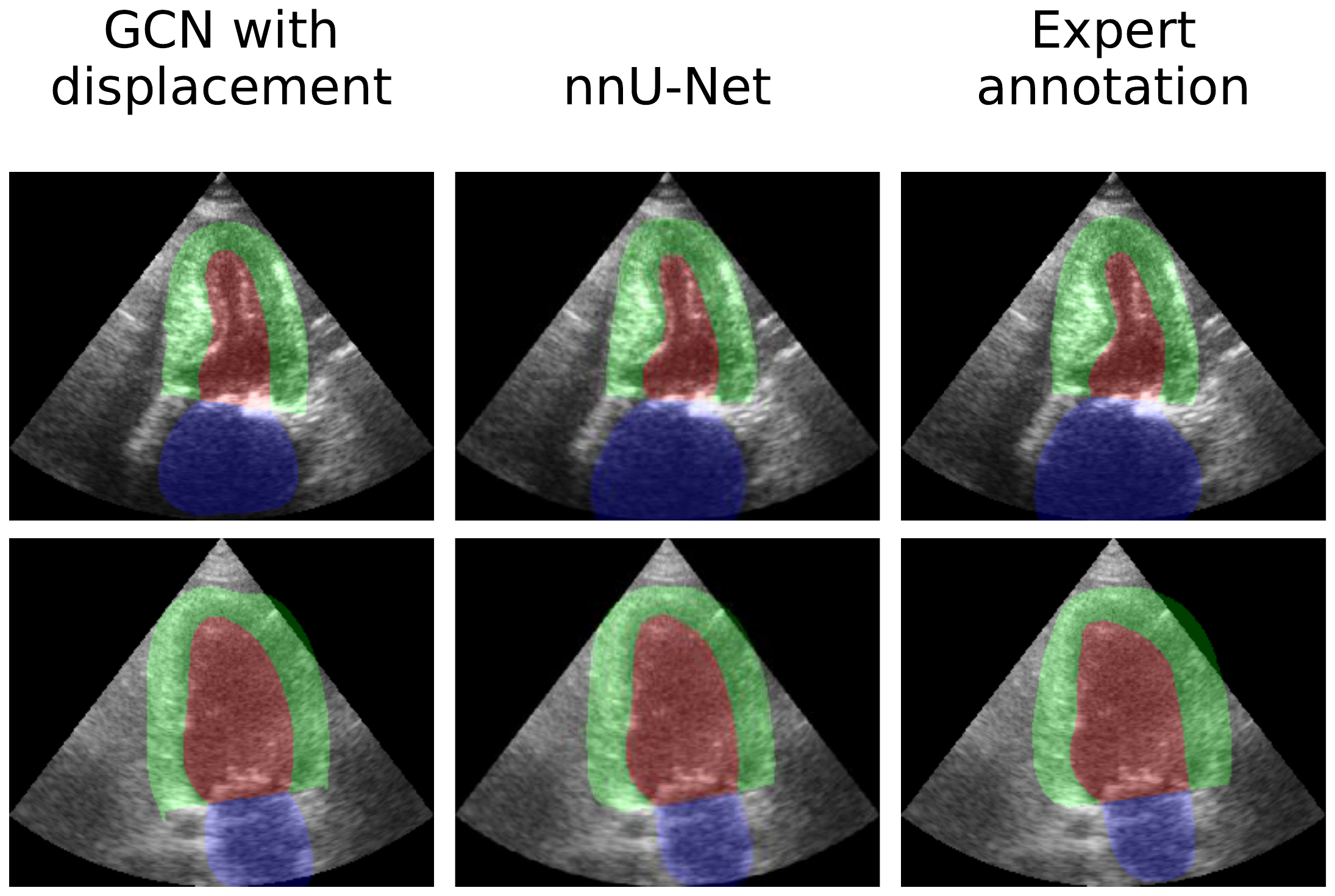}
   \caption{Median cases}
   \label{fig:median} 
\end{subfigure}

\begin{subfigure}[b]{0.8\textwidth}
   \includegraphics[trim={0cm 0cm 0cm 0cm},clip,width=1\linewidth]{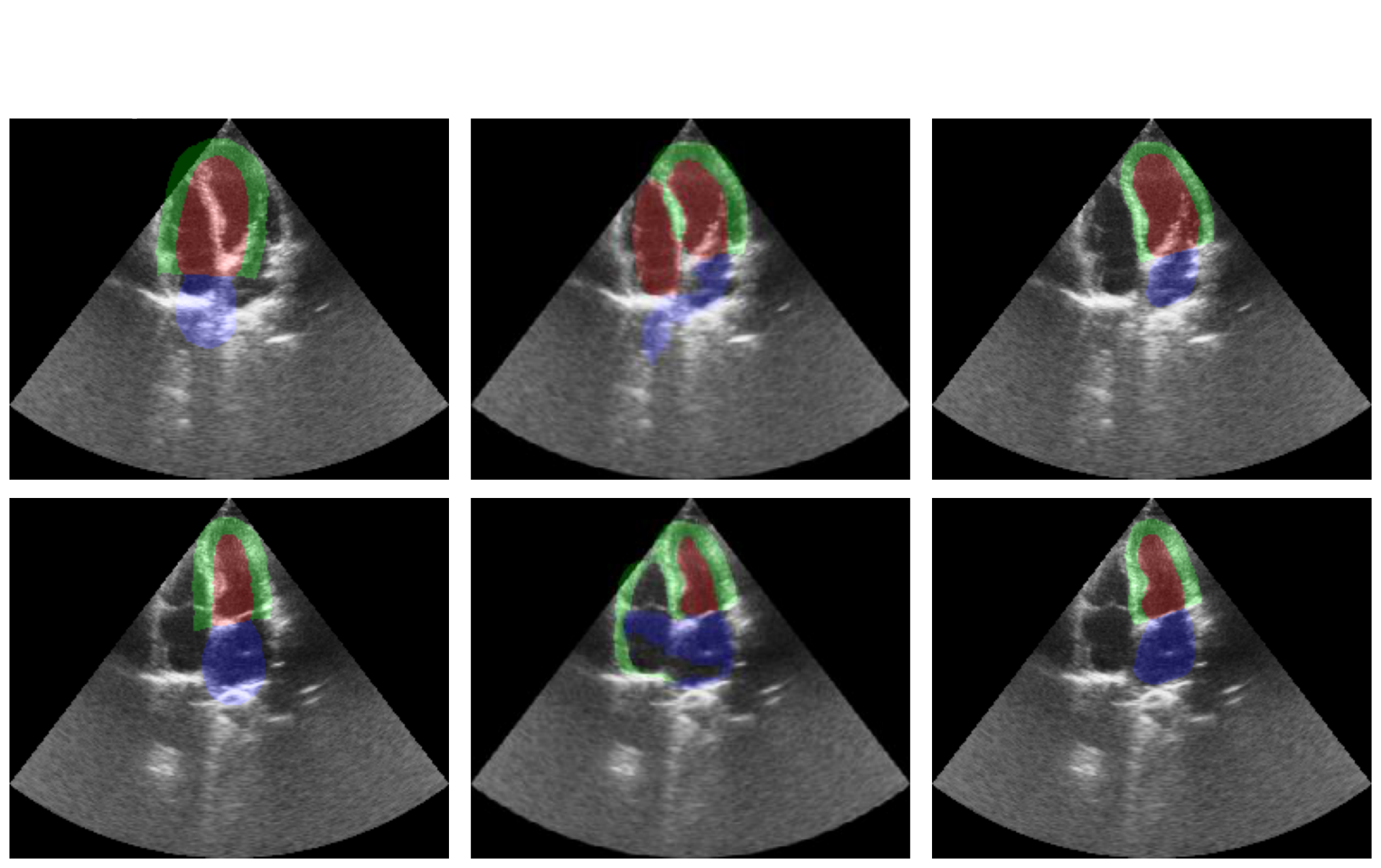}
   \caption{Worst cases}
   \label{fig:worst}
\end{subfigure}
\caption{Case analysis and comparison of the GCN with displacement method and nnU-Net on CAMUS. The cases are selected based on the overall Dice score between the annotation and the GCN or U-Net segmentations.}
\label{fig:cases}
\end{figure*}

 The anatomically incorrect cases for the GCN are cases where the model confuses the endo- and epicardium keypoints, as in Fig.~\ref{fig: ring_error}. Using the displacement method avoids this from happening, resulting in no anatomically invalid segmentations. Fig.~\ref{fig:cases} shows the median and worst cases of the GCN with the displacement method and nnU-Net.


\subsection{Inter-model agreement as quality predictor}

\begin{figure}
\centering
  \centering
  \includegraphics[width = 0.5\linewidth]{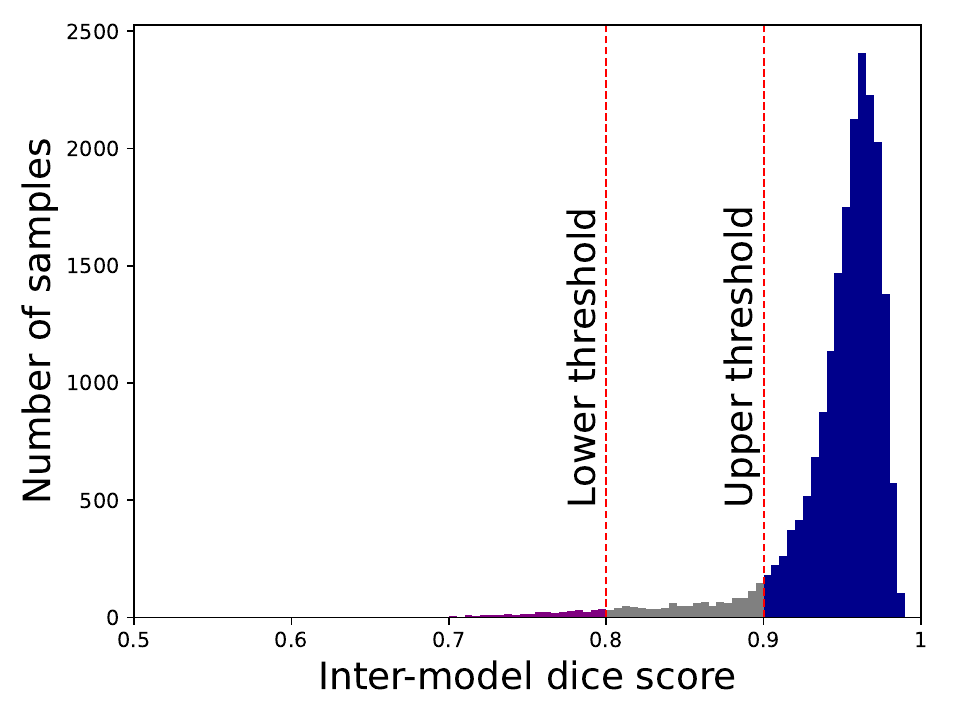}
  \caption{Histogram of inter-model Dice scores between nnU-Net and GCN on the HUNT4 evaluation set. The upper and lower threshold determine the set of images that are sampled for the experiment that evaluates inter-model agreement as a quality predictor.}
  \label{fig: hist}
\end{figure}

The goal of this experiment was to evaluate the proposed inter-model agreement as a segmentation quality predictor and to investigate whether GCN has a better worst-case performance than nnU-Net.
The inter-model agreement is measured as the Dice similarity between the segmentation masks of nnU-Net and GCN with displacement method. For this experiment, we use the clinical dataset HUNT4 as it better represents the ultrasound images used in practice. Fig.~\ref{fig: hist} shows a histogram of the inter-model Dice scores on the HUNT4 evaluation set. For the experiment, the samples are split between cases with high and low inter-model agreement. The threshold for low inter-model Dice is 0.8, equalling ${\sim}2\%$ of cases. The threshold of high inter-model dice is 0.9, equalling ${\sim}92\%$ of cases. A total of 200 samples were extracted by combining and randomizing 100 samples below the low threshold and 100 samples above the high threshold. Each sample contains 3 images: the input ultrasound image and the two segmentations produced by the two architectures in random order. A clinician manually classified each sample using the following criteria:
\begin{itemize}
  \item Overall quality of ultrasound image: 'High', 'Medium', 'Low', or '\textbf{Unsuitable} for measurements'. The last category is reserved for images with so low quality that they should not be used for any clinical measurement. 
  \item Correct view?: 'Yes' or 'No'. 'Yes' means the view is A4C or A2C.
  \item LV-focused?: 'Yes' or 'No'. 'Yes' means the LV is central in the image and the depth is adjusted to the LV.
  \item Anatomically correct segmentation? 'Yes' or 'No'. We use the same criteria for anatomical correctness as Painchaud et al. \cite{painchaud2020cardiac}.
  \item  Correct placement of segmentation? 'Yes' or 'No'. 'Yes' means the segmentation is placed correctly.
\end{itemize}

\begin{figure}
\centering
  \centering
  \includegraphics[width = 0.5\linewidth]{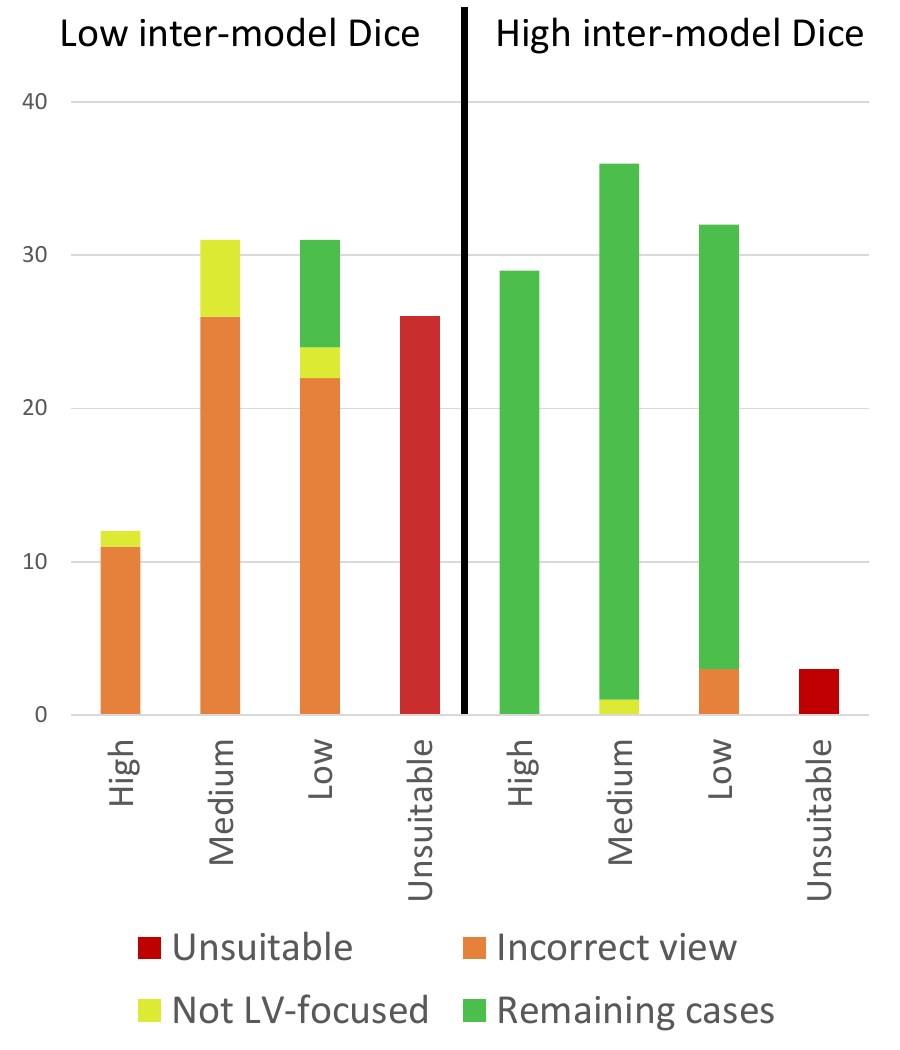}
  \caption{Comparison of input quality between samples with a high and a low inter-model agreement. The exact number of cases can be found in Table \ref{table: quality experiment}. From the 100 high agreement samples, 93 were in distribution, while from the 100
low agreement samples, only 7 were in distribution.}
  \label{fig: inter-model quality}
\end{figure}

Table \ref{table: quality experiment} summarizes the results of the experiment in which the first subtable (a) shows that when the inter-model agreement is high (> 0.9) the two networks both produce valid segmentations. The second subtable (b) shows that when the inter-model agreement is low (< 0.8), the GCN produces slightly more correct segmentations. If we regard samples that have suitable image quality, the correct view, and are LV-focused as in-distribution samples, we see that from the 100 high agreement samples, 93 were in distribution, while for the 100 low agreement samples, only 7 were in distribution. Furthermore, all 7 in-distribution samples with low agreement had low quality.
Fig.~\ref{fig: inter-model quality} visualizes these results showing the difference in input quality between samples with high and low inter-model agreement.

\begin{table} 
\scriptsize
\begin{center}
\caption{Results of the experiment on using the inter-model agreement as a quality predictor. The GCN uses the displacement method in this experiment.}
\begin{subtable}{1\linewidth}
\caption{Analysis for cases with high inter-model agreement}
\setlength{\tabcolsep}{3pt}
\begin{tabular}{m{80pt} m{80pt}m{60pt}|m{50pt}m{50pt}|m{50pt}m{50pt}}
\toprule
\textbf{Image set} & \raggedright \textbf{Quality of ultrasound image} & \raggedright \textbf{Number of cases} & \multicolumn{2}{m{100pt}|}{\raggedright  \textbf{Number of anatomically correct cases}} & \multicolumn{2}{m{100pt}}{\raggedright \textbf{Number of cases with correct placement }}\\
\cline{4-7}

 &  & & \vspace{0.1cm} nnU-Net  & \vspace{0.1cm}  GCN & \vspace{0.1cm}  nnU-Net  & \vspace{0.1cm}  GCN \\
\midrule
 & High & 29 & 29 & 29 & 29 & 29\\
All  & Medium & 36 & 36 & 36  & 35 &35\\
images & Low & 32 & 31 & \textbf{32}  & 30 & \textbf{31}\\
 & Unsuitable & 3 & - & - &- &- \\
\hline
\multirow{3}{1.4cm}{Correct view}& High & 29 & 29 & 29 &29 & 29\\
& Medium & 36 & 35 &35 &34 & 34\\
 & Low & 29 & \textbf{27} & 26& \textbf{26} & 25\\
\hline
\multirow{3}{2.5cm}{Correct view and LV-focused} & High & 29 & 29 & 29& 29 & 29\\
  & Medium & 35 &35 &35& 34 & 34 \\
 & Low & 29 &\textbf{26} &25&\textbf{26} & 25\\
\bottomrule
\end{tabular}
\end{subtable}
\begin{subtable}{1\linewidth}
\vspace{20pt}
\caption{Analysis for cases with low inter-model agreement}
\setlength{\tabcolsep}{3pt}
\begin{tabular}{m{80pt} m{80pt}m{60pt}|m{50pt}m{50pt}|m{50pt}m{50pt}}
\toprule
\textbf{Image set} & \raggedright \textbf{Quality of ultrasound image} & \raggedright \textbf{Number of cases} & \multicolumn{2}{m{100pt}|}{\raggedright  \textbf{Number of anatomically correct cases}} & \multicolumn{2}{m{100pt}}{\raggedright \textbf{Number of cases with correct placement }}\\
\cline{4-7}

 &  & & \vspace{0.1cm} nnU-Net  & \vspace{0.1cm}  GCN & \vspace{0.1cm}  nnU-Net  & \vspace{0.1cm}  GCN \\
\midrule
 & High & 12 & 12 &  12 & 0 & 0 \\
All  & Medium & 31  & 26 & \textbf{29}  & 0 & \textbf{2}\\
images & Low & 31 & 25 & \textbf{30} & 2 & \textbf{4}\\
 & Unsuitable & 26 & - & - &- &- \\
\hline
\multirow{3}{1.4cm}{Correct view}& High &  1& 1  & 1 & 0& 0\\
& Medium & 5 & 2 & \textbf{5}& 0& \textbf{2}\\
 & Low & 9 & 7 & \textbf{9} & 2 & 2\\
\hline
\multirow{3}{2.5cm}{Correct view and LV-focused} & High &  0 & - & - & - & -\\
  & Medium & 0 & - & - & - & - \\
 & Low & 7 & 7 & 7&2 & 2\\
\bottomrule

\end{tabular}
\end{subtable}
\label{table: quality experiment}

\end{center}

\end{table}

\subsection{Clinical evaluation on HUNT4}

We use the clinical dataset HUNT4 to evaluate the accuracy of clinical measurements instead of CAMUS because the CAMUS dataset only contains volume measurements which were measured using the annotations and not using clinically approved software. The HUNT4 dataset also contains considerably more patients.
We compare the accuracy of the U-Net and the GCN which both were trained on the HUNT4 training set.
The GCN used for this experiment is the GCN with displacement method. 

The automatic estimation of EF mimics the procedure for manual LV volume calculations 
suggested by clinical guidelines \cite{lang2015recommendations}. The automatic procedure contains the following steps:

\begin{enumerate}
\item Detect ED and ES of each cycle using the timing neural network described by Fiorito et al. \cite{fiorito2018detection} for both the A2C and A4C recordings of the same patient. This means the view was labeled manually during acquisition and the timing is estimated using deep learning.
\item Segment the ED and ES frame of each cycle for both the A2C and A4C recordings of the same patient.
\item Combine the \textit{usable} segmentations in the A2C recordings with the \textit{usable} segmentations in the A4C recordings for biplane volume calculations using the modified Simpson method. A segmentation is \textit{unusable} if the procedure described in section \ref{subsec: data preproc} fails to extract the LV apex or base points, which position is required for the modified Simpson method. The EF is then calculated as $EF=\frac{{ED\:volume}-ES\:volume}{ED\:volume}$. Measurements from all cardiac cycles are combined and averaged.
\end{enumerate}

Our results include 1877 out of 1913 patients, equalling a feasibility of 98.1\%. In 13 cases the timing network failed. In the remaining 23 cases, either U-Net 1 or nnU-Net failed to produce a \textit{usable} segmentation for all cycles. Because the GCN predicts the LV apex and base points directly as one of the keypoints, it can produce an EF for every recording. However, for a fair comparison, we exclude these cases from our results.

\begin{figure*}
     \centering
     \begin{subfigure}{0.32\linewidth}
         \centering
         \includegraphics[trim={0.9cm 1.6cm 0.4cm 2cm},clip,width=\textwidth]{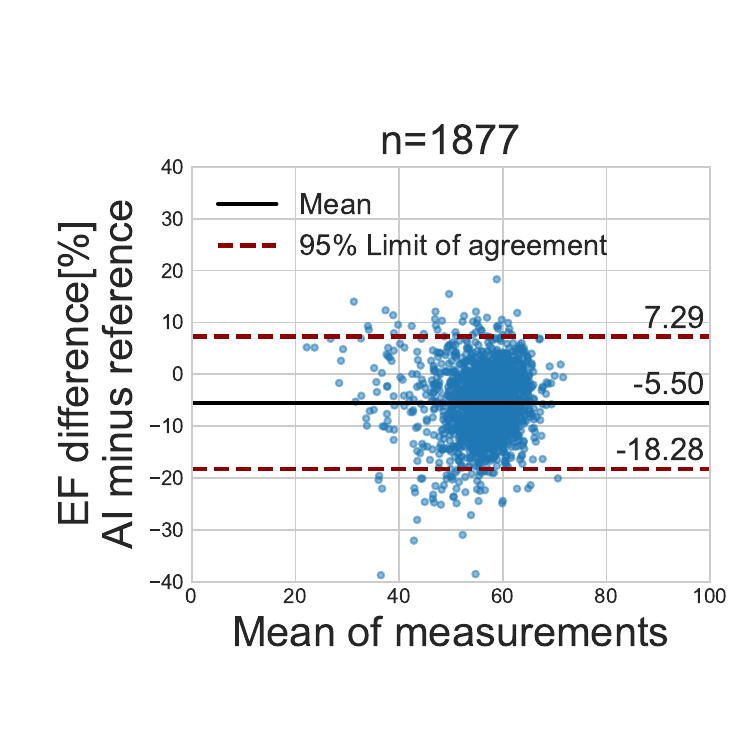}
         \caption{U-Net 1}
     \end{subfigure}
     \hfill
     \begin{subfigure}[b]{0.32\textwidth}
         \centering
         \includegraphics[trim={0.9cm 1.6cm 0.4cm 2cm},clip, width=\textwidth]{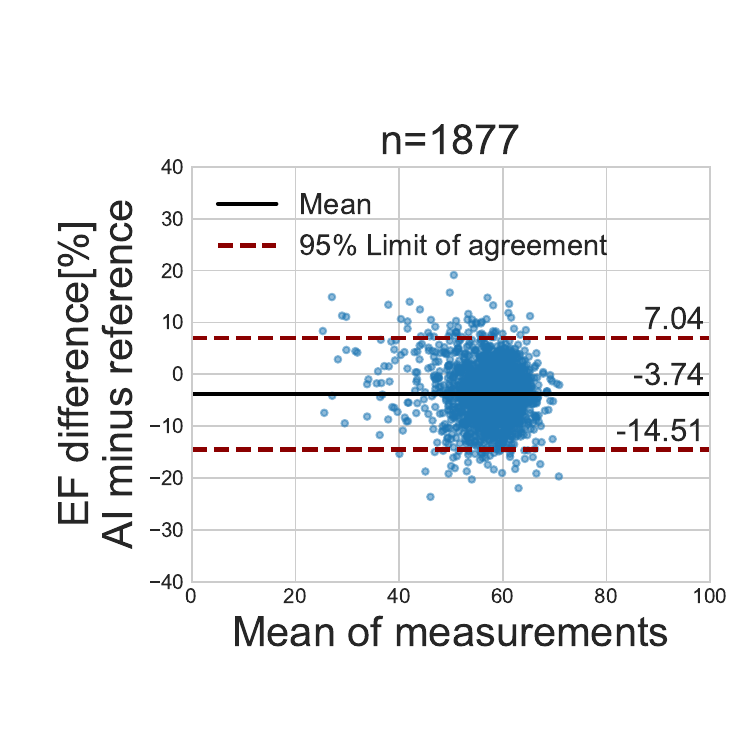}
         \caption{nnU-Net}
     \end{subfigure}
    \hfill
     \begin{subfigure}[b]{0.32\textwidth}
         \centering
         \includegraphics[trim={0.9cm 1.6cm 0.4cm 2cm},clip, width=\textwidth]{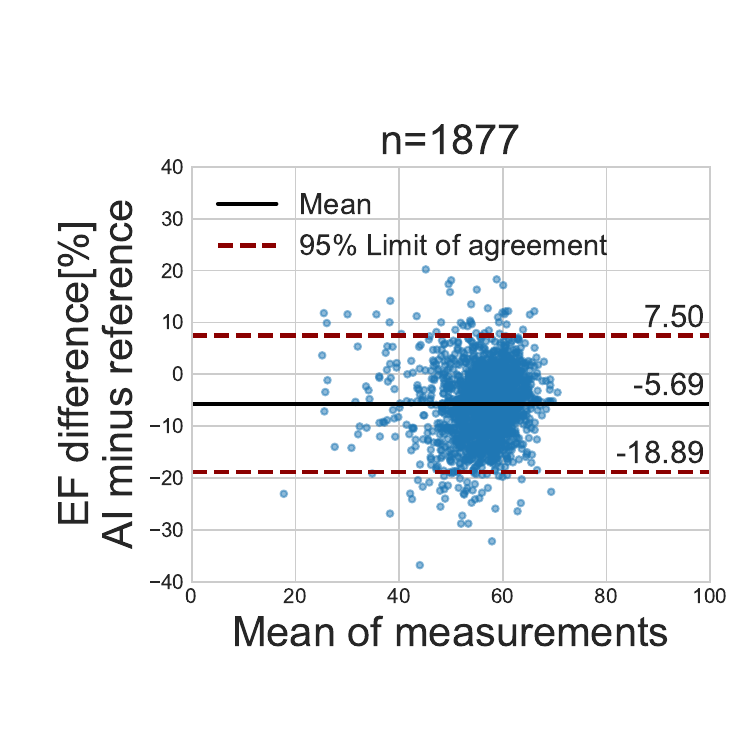}
         \caption{GCN}
     \end{subfigure}
        \caption{Bland–Altman 
        plots of the automatic EF measurements compared to the manual reference. All frames with a usable segmentation are used.
        }
        \label{fig: ef}
\end{figure*}

Fig.~\ref{fig: ef} shows the Bland-Altman plots of the automatic methods compared to the manual reference.
Table \ref{table: eval ef} shows the Mean Absolute Error (MAE) of EF measurements between the automatic method and manual reference. The ensembles average the EF estimates of the two models. The difference in MAE between nnU-Net and all other methods is significant with $p<0.05$ using the Wilcoxon signed-rank test\cite{woolson2007wilcoxon}.

\begin{figure*}
     \centering
     \begin{subfigure}{0.32\linewidth}
         \centering
         \includegraphics[trim={0.9cm 1.6cm 0.4cm 2cm},clip,width=\textwidth]{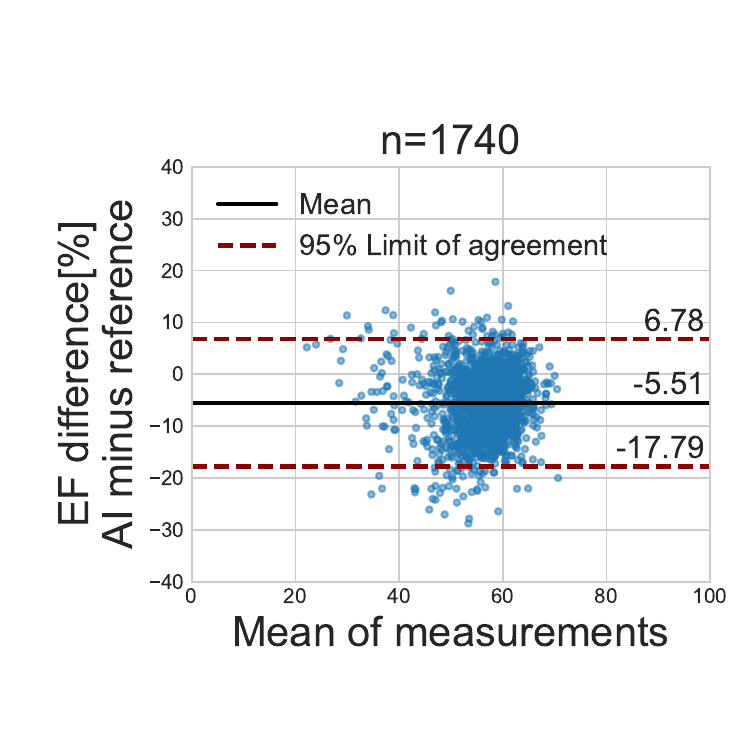}
         \caption{U-Net 1}
     \end{subfigure}
     \hfill
     \begin{subfigure}[b]{0.32\textwidth}
         \centering
         \includegraphics[trim={0.9cm 1.6cm 0.4cm 2cm},clip, width=\textwidth]{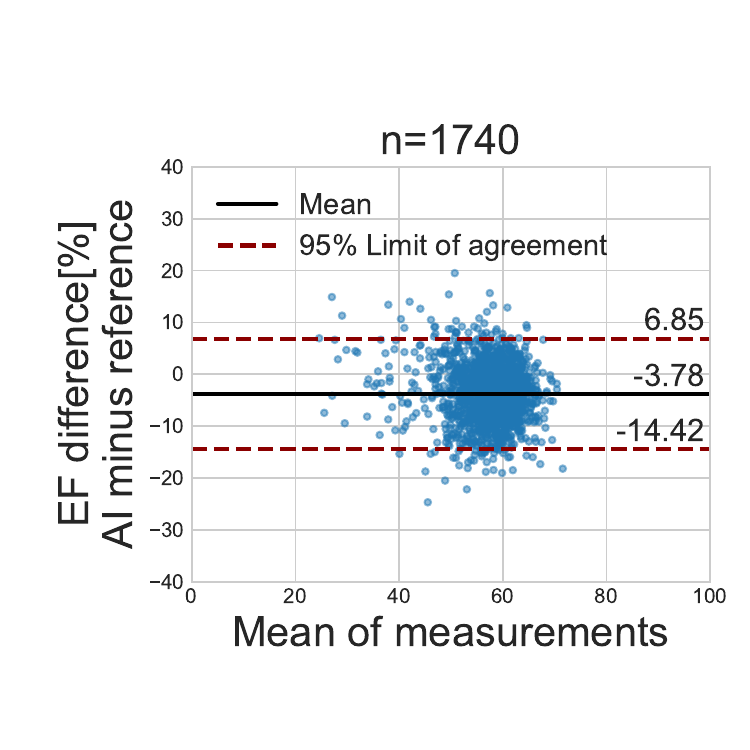}
         \caption{nnU-Net}
     \end{subfigure}
    \hfill
     \begin{subfigure}[b]{0.32\textwidth}
         \centering
         \includegraphics[trim={0.9cm 1.6cm 0.4cm 2cm},clip, width=\textwidth]{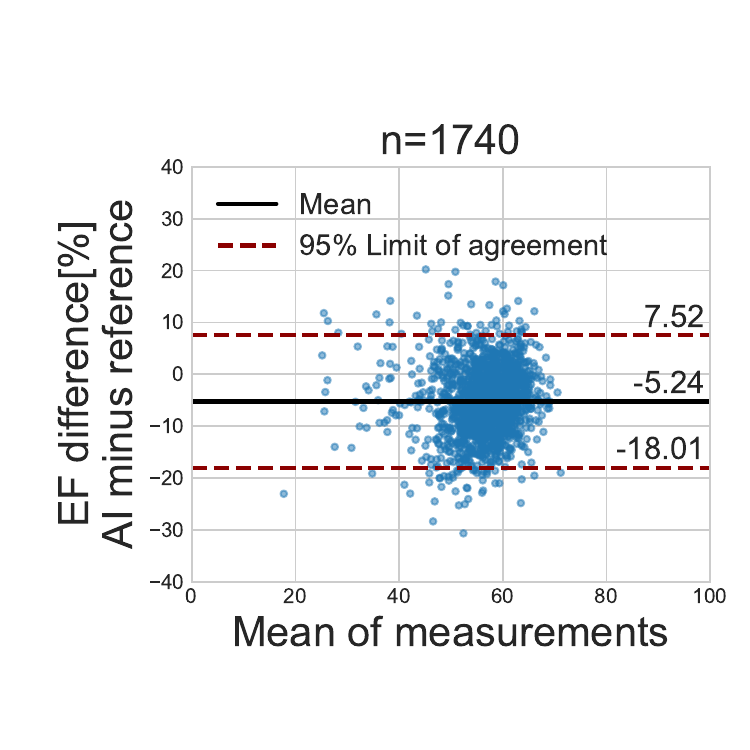}
         \caption{GCN}
     \end{subfigure}
        \caption{Bland–Altman 
        plots of the automatic EF measurements compared to the manual reference when only using the frames for which the inter-model Dice agreement is more than 0.85.
        }
        \label{fig: ef filtered}
\end{figure*}

Finally, we use the findings of the previous experiment and filter the ED and ES frames based on inter-model agreement. Fig.~\ref{fig: ef filtered} shows the same Bland-Altman plots when only using the frames where the inter-model Dice between U-Net or nnU-Net and the GCN is above 0.85. In 137 out of 1877 cases, there was no ED or ES frame in all three cycles of the recording with Dice above 0.85, resulting in the patient being excluded. As a result, the filtered results only include 1740 out of 1913 patients, equalling a feasibility of 91\%.

\begin{table}
\begin{center}
\scriptsize
\caption{Evaluation of automatic EF estimation on HUNT4. The mean absolute error of nnU-Net is statistically significantly lower than all the other models.}
\label{table: eval ef}
\begin{tabular}{l l}
\toprule
\textbf{Method} & \textbf{Mean absolute error} \\
\midrule 
U-Net 1  \cite{leclerc2019deep} & 6.79$\pm$6.5 \\
nnU-Net  \cite{isensee2021nnu} & \textbf{5.38$\pm$5.5}\\
GCN with displacement & 7.07$\pm$6.7 \\
U-Net 1 GCN with displacement ensemble & 6.66$\pm$6.1 \\
nnU-Net U-Net 1 ensemble & 5.92$\pm$5.7\\
nnU-Net GCN with displacement ensemble & 6.00$\pm$5.7 \\
\bottomrule
\end{tabular}
\label{tab1}
\end{center}
\end{table}

\subsection{Inference time}
Table \ref{table: runtime} summarizes the inference times of the proposed architectures and baseline models on an NVIDIA GeForce RTX  3070 Ti Laptop GPU. To measure the inference time, the model performs a warmup run on 1000 random dummy inputs. Afterward, the model performs 10 test runs on 100 random inputs. The final inference time is the average runtime of each of the test runs divided by 100. These values do not include the data loading time and pre- and post-processing steps.

\begin{table} 
\begin{center}
\scriptsize
\caption{Inference time comparison of the proposed architectures and baseline models on an NVIDIA GeForce RTX  3070 Ti Laptop GPU.}
\label{table: runtime}
\setlength{\tabcolsep}{3pt}
\begin{tabular}{m{180pt}m{100pt}m{120pt}}
\hline
Architecture & Inference time in ms & Number of parameters in millions \\
\hline

nnU-Net \cite{isensee2021nnu} & 5.40$\pm$0.008 & 33.4\\
U-Net 1 \cite{leclerc2019deep}& 1.88$\pm$0.011 & 2.0\\
MobileNet-v2-GCN & \textbf{1.35$\pm$0.007}&3.3\\
MobileNet-v2-GCN with displacement method& 1.40$\pm$0.020&3.3\\
ResNet-50-GCN & 3.37$\pm$0.010 &25.2\\
ResNet-50-GCN with displacement method& 3.55$\pm$0.009&25.2 \\
\hline
\end{tabular}
\label{tab1}
\end{center}
\end{table}

\section{Discussion}
\label{sec: discussion}

\subsection{Anatomical correctness and segmentation accuracy}

In this work, we developed a multi-structure graph convolutional network (GCN) for cardiac ultrasound segmentation.
While the GCN method can eliminate anatomical incorrect segmentations, it comes at the cost of slightly reducing the segmentation accuracy compared to U-Net.
Whereas the GCN does not outperform U-Net in terms of accuracy for multi-structure segmentation, the keypoint representation has its own advantages. Both architectures have their unique data representation and solve the problem in a fundamentally different way. The pixel-wise approach of U-Net can give the most accurate segmentations, but can also produce anatomically incorrect results. On the other hand, the GCN with displacement method has lower accuracy but does not produce anatomically incorrect segmentations as the cardiac shapes are embedded as a strong bias in the network. However, the GCN can still fail to place the anatomically correct shape correctly in the image.
Fig.~\ref{fig:worst} shows an example where both the GCN and U-Net fail to generalize to samples with increased depth. From an abstract point of view, GCN can be seen as a heavily regularised version of pixel-wise architectures. The displacement method regularizes the architecture even further. Finally, it is worth noting that nnU-Net is an order of magnitude larger and slower than the GCN.

\subsection{Ablation study}

The results of the ablation study show that reducing the depth of the decoder changes the performance of the GCN non-significantly, indicating that the pre-trained CNN is the most important contributor. There are two possible explanations for this behavior. It could be that the architecture of the GCN decoder is not efficient for processing data in keypoint embedding space. 
On the other hand, it could also be that the decoder part is superfluous, as the shapes of the cardiac structures are easy enough to not require a shape encoding as the embedding of the graph convolutional layers.

\subsection{Combination with U-Net}


Combining GCN and U-Net in a cascade gives a mixed result of the characteristics of both architectures. For the GCN - nnU-Net cascade, there are no additional augmentations to the GCN output that serve as an extra input channel to the nnU-Net, as it is used out of the box. As a result, the nnU-Net cascade relies too much on the GCN output and only learns to do minor adjustments to the GCN intermediates. On the other hand, the GCN - U-Net 1 training procedure augments the GCN intermediates with additional rotation, scaling, and translation augmentations specifically for the cascade network, as described in subsection \ref{subsec: cascade}. This results in the U-Net correcting some cases of wrong placement of the GCN but also re-introduces anatomical incorrect outliers. Table \ref{table: comp CAMUS} reflects these findings numerically.

\subsection{Inter-model agreement}
Despite the excellent Dice and Hausdorff metrics of U-Net, this pixel-wise segmentation approach can fail completely in certain cases as shown in Fig.~\ref{fig:anatomical_incorrect_example}. In this work, we have proposed using the inter-model agreement as a simple and effective approach to detect these failing cases.

The distinct characteristics of U-Net and GCN make them appealing for joint use to quantify uncertainty. Although averaging their clinical estimates doesn't enhance accuracy, the agreement between the models offers valuable insight for identifying difficult cases and erroneous segmentations of U-Net. The leading approach to uncertainty quantification in deep learning is Bayesian Neural Networks \cite{izmailov2021bayesian}, but their multiple passes during inference hinder real-time use. Deep ensembles are another option \cite{izmailov2021bayesian}, but require multiple models to run on each input. Combining U-Net with GCN approximates deep ensembles practically. Specifically, U-Net and GCN are unlikely to make the same mistake since the GCN is designed to avoid the most common error of U-Net i.e. anatomically incorrect segmentations.

The results of the experiments using inter-model agreement indicate a clear distinction in input quality between high and low agreement cases. Fig.~\ref{fig: inter-model quality} shows significantly fewer high-quality images and fewer unsuitable cases with high inter-model agreement as compared to the cases with low inter-model agreement ($p<0.05$). Furthermore, most of the cases with low inter-model agreements were the wrong view. In only 15 out of 74 suitable low agreement cases the view was correct. The majority (55) of these wrong view cases are apical long axis (ALAX) views or views rotated towards ALAX. These images are in the dataset because the clinicians used these views in practice even though they were instructed to use A4C and A2C views to measure EF. One of the reasons is that sometimes it is not possible to get a clear A2C or A4C view so the clinician trades view correctness for image quality. However, the segmentation models are only trained on LV-focused A2C and A4C views, resulting in poor performance. Finally, only 7 out of 100 low agreement cases had suitable quality, the right view, and were LV-focused as compared to 93 out of 100 high agreement cases. This demonstrates the inter-model agreement estimator as an efficient method to detect out-of-distribution cases in the form of very low image quality or wrong views. 

\subsection{Clinical evaluation}

For accuracy on EF measurements, the nnU-Net exhibits the narrowest limits of agreement and the fewest outliers in comparison to both GCN and U-Net 1. However, when frames with low inter-model agreement are omitted, the limits of agreement for all three networks improve, as can be observed when comparing Figs.~\ref{fig: ef} and \ref{fig: ef filtered}. This improvement can be attributed to the elimination of outliers stemming from problematic inputs or flawed segmentations. Thus, while the GCN does not improve EF accuracy directly, it can help discard cases where automatic EF is not appropriate, consequently narrowing the limits of agreements for the remaining cases. The negative mean in Figs.~\ref{fig: ef} and \ref{fig: ef filtered} show that each of the deep learning methods underestimates the reference ejection fraction. Because there are many possible sources of error, both from the deep learning methods and from the reference annotations, the reason for the negative bias is not trivial. The article by Olaisen et al \cite{Olaisen2023-fy} discusses this more in-depth.


\section{Real-time demo application}
\begin{figure}
\centering
  \centering
  \includegraphics[width = 0.5\linewidth]{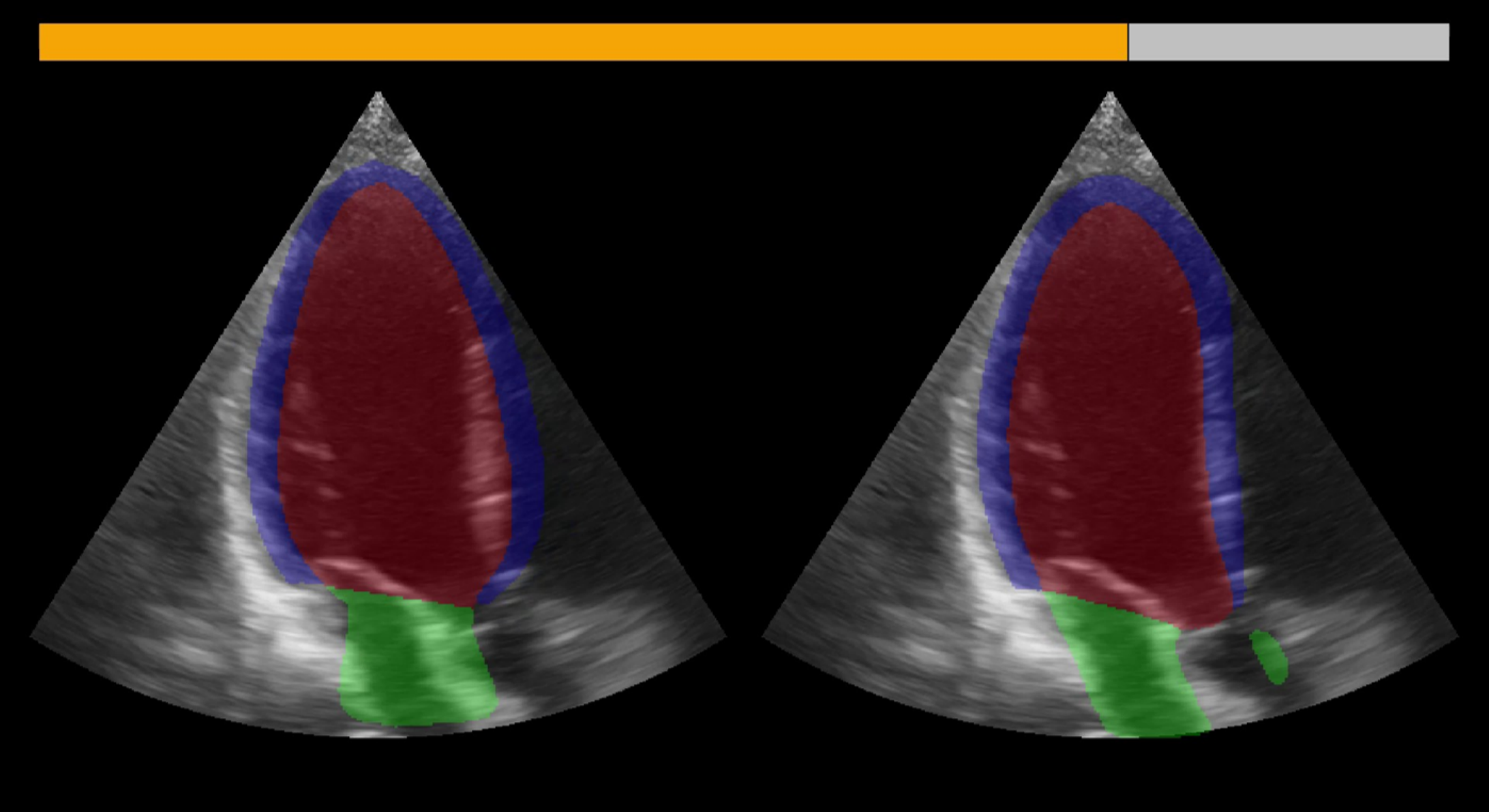}
  \caption{Screenshot of the real-time demo application. The GCN and nnU-Net segmentations are shown on the left and right side respectively. The color-coded status bar on top visualizes the agreement between the models.}
  \label{fig: screenshot demo}
\end{figure}

To demonstrate the potential of the inter-model agreement to detect out-of-distribution cases and faulty segmentations in real-time, a real-time application was created using the FAST framework \cite{8844665}. It shows the segmentation output of the GCN and nnU-Net side by side together with a status bar that visualizes the agreement between the models. Fig.~\ref{fig: screenshot demo} shows a screenshot of the application in action. We created a demo video \cite{VanDeVyver2023} that shows the application in use while a clinician is operating a GE Vivid E95 scanner. The video demonstrates the effectiveness of the inter-model agreement as a method to detect out-of-distribution and low image quality cases. The video is available at \url{https://doi.org/10.6084/m9.figshare.24230194}.

\section{Conclusion}
\label{sec: conclusion}

This paper proposes a multi-structure graph convolutional network (GCN) to reduce anatomical incorrect results produced by U-Nets in cardiac ultrasound segmentation. We propose the displacement method as a clinically motivated regularization that eliminates anatomically incorrect segmentations for multi-structure segmentation. The ablation study showed that changing the architecture of the GCN decoder does not affect performance significantly. This indicates the encoder is the most important component of the network

Measuring the inter-model agreement between GCN and U-Net yields a simple and highly effective method for quantifying uncertainty and thereby detecting out-of-distribution and failing cases. Whereas the average performance of both U-Net and the GCN is within inter-observer variability, their worst-case behavior is still drastically worse than human annotators. These outliers occur mostly due to a lack of generalization capabilities, which is still a fundamental problem in cardiac segmentation.
 Low inter-model agreement can detect out-of-distribution cases and thus can be used as a warning signal to the user using the automatic segmentation tool in practice. 
 
 The idea of GCNs is a relatively new concept for doing cardiac segmentation that shows promising initial results towards more robust cardiac segmentation. The GCN can be implemented as a robust independent model or as a supplementary model to validate the segmentation outcomes of pixel-wise techniques such as U-Net.


\bibliographystyle{IEEEtran}

\bibliography{bibliography.bib}
\end{document}